\documentclass[runningheads]{llncs}

\usepackage[all]{xy}
\usepackage[dvips]{graphicx}
\usepackage{amssymb}
\usepackage[mathcal]{euscript}
\usepackage{proof}

\usepackage{syntonly}
\usepackage[american]{babel}

\newcommand{\define}[1]{\textit{#1}}

\newcommand{\qstates}{\ensuremath{\mathbb{S}}}
\newcommand{\vstates}{\ensuremath{\mathbb{V}}}

\newcommand{\Variableterms}{\ensuremath{\mathcal{V}_{{\it term}}}}

\newcommand{\UnitaryGates}[1]{\ensuremath{\mathcal{U}^{#1}}}
\newcommand{\FV}{\ensuremath{{\it FV}}}

\newcommand{\bangnoted}{\mbox{exponential}}
\newcommand{\nbang}[2]{{!^{#1}{#2}}}
\newcommand{\bang}[1]{{!{#1}}}

\newcommand{\probred}[1]{\mathbin{\longrightarrow_{#1}}}
\newcommand{\nprobred}[2]{\mathbin{\longrightarrow_{#2}^{#1}}}
\newcommand{\probreach}{\mathbin{\longrightarrow}}
\newcommand{\nprobreach}{\mathbin{\longrightarrow^*}}

\newcommand{\Rreach}{\mathbin{\rightsquigarrow}}
\newcommand{\nRreach}{\mathbin{\rightsquigarrow^*}}

\newcommand{\CBV}{\mathbin{\longrightarrow_{CBV}}}

\newcommand{\loli}{\mathbin{\multimap}}
\newcommand{\duploli}{\mathbin{\Rightarrow}}

\newcommand{\subtype}{\mathbin{<\! :}}
\newcommand{\entail}{\mathbin{{\vartriangleright}}}
\newcommand{\skelentail}{\mathrel{{\blacktriangleright}}}

\newcommand{\skel}[1]{{{}^{\dag}#1}}

\newcommand{\void}[1]{}

\newcommand{\iftermx}[3]{\mathop{{\it if}}#1\mathbin{{\it then}}#2\mathbin{{\it else}}#3}

\newcommand{\newterm}{\mathop{\it new}}
\newcommand{\measureterm}{\mathop{\it meas}}
\newcommand{\prodterm}[1]{{\langle}#1{\rangle}}
\newcommand{\produnitterm}{\mathop{\ast}}
\newcommand{\letprodterm}[3]{{\it let}\;\prodterm{#1}{=}#2\;{\it in}\;#3}

\newcommand{\bittype}{\mathop{{\it bit}}}
\newcommand{\qbittype}{\mathop{{\it qbit}}}
\newcommand{\produnittype}{\top}

\newcommand{\ket}[1]{|#1\rangle}
\newcommand{\s}[1]{\{#1\}}
\newcommand{\qbit}{{\it qbit}}
\newcommand{\bit}{{\it bit}}
\newcommand{\binrep}[2]{{\ulcorner{#1}\urcorner^{#2}}}

\newcommand{\invskel}[1]{{ ^{\clubsuit}#1}}
\newcommand{\skeltotype}{\looparrowright}

\newcommand{\abs}[1]{|#1|}
\newcommand{\iso}{\cong}
\renewcommand{\leq}{\leqslant}
\renewcommand{\geq}{\geqslant}
\newcommand{\bor}{\;|\;}
\newcommand{\such}{~|~}
\spnewtheorem{convention}{Convention}{\itshape}{\rmfamily}
\newcommand{\qType}{{\it qType}}
\newcommand{\iType}{{\it iType}}
\newcommand{\ii}{\rightarrow}
\newcommand{\typ}[2]{#1{:}#2}
\newcommand{\uniq}[1]{{}^{\#}\!#1}

\newenvironment{proofof}[1]{\paragraph{{\it Proof of #1.}}}{\par}

\begin{document}

\mainmatter 

\title{A lambda calculus for quantum
  computation with classical control}
\titlerunning{Functional Programming for Quantum Computation}

\author{Peter Selinger, Beno{\^\i}t Valiron}
\authorrunning{P. Selinger, B. Valiron}

\institute{University of Ottawa}
\date{Oct. 29, 2004}

\maketitle

\begin{abstract}
  
  The objective of this paper is to develop a functional programming
  language for quantum computers. We develop a lambda calculus for the
  classical control model, following the first author's work on
  quantum flow-charts. We define a call-by-value operational
  semantics, and we give a type system using affine intuitionistic
  linear logic. The main results of this paper are the safety
  properties of the language and the development of a type inference
  algorithm.

\end{abstract}

\section{Introduction}

The objective of this paper is to develop a functional programming
language for quantum computers. Quantum computing is a theory of
computation based on the laws of quantum physics, rather than of
classical physics.  Quantum computing has become a fast growing
research area in recent years. For a good introduction, see e.g.
{\cite{nielsen02,preskill99}}.

Due to the laws of quantum physics, there are only two kinds of basic
operations that one can perform on a quantum state, namely {\em
  unitary transformations} and {\em measurements}. Many existing
formalisms for quantum computation put an emphasis on the former,
i.e., a computation is understood as the evolution of a quantum state
by means of unitary gates. Measurements are usually performed at the
end of the computation, and outside of the formalism. In these models,
a quantum computer is considered as a purely quantum system, i.e.,
without any classical parts. One example of such a model is the
quantum Turing machine {\cite{benioff80,deutsch85}}, where the
entire machine state, including the tape, the finite control, and the
position of the head, is assumed to be in quantum superposition.
Another example is the quantum lambda calculus of van
Tonder~\cite{tonder03,tonder04}, which is a higher-order, purely
quantum language without an explicit measurement operation.

On the other hand, one might imagine a model of a quantum computer
where unitary operations and measurements can be interleaved. One
example is the so-called {\em QRAM model} of Knill~\cite{knill96},
which is also described by Bettelli, Calarco and
Serafini~\cite{bettelli03}. Here, a quantum computer consists of a
classical computer connected to a quantum device. In this
configuration, the operation of the machine is controlled by a
classical program which emits a sequence of instructions to the
quantum device for performing measurements and unitary operations.  In
such a model, the control structures of the machine are classical, and
only the data being operated upon is quantum.  This situation is
summarized by the slogan ``quantum data, classical control''
{\cite{selinger04}}. Several programming languages have been proposed
to deal with such a model {\cite{bettelli03,sanders00}}. The present
paper is based on the work of \cite{selinger04}.

In this paper, we propose a {\em higher-order} quantum programming
language, i.e., one in which functions can be considered as data. In
our language, a program is a lambda term, possibly with some quantum
data embedded inside. The basic idea is that lambda terms encode the
control structure of a program, and thus, they would be implemented
classically, i.e., on the classical device of the QRAM machine.
However, the data on which the lambda terms act is possibly quantum,
and is stored on the QRAM quantum device.

Because our language combines classical and quantum features, it is
natural to consider two distinct basic data types: a type of
\define{classical bits} and a type of \define{quantum bits}. They
behave in a complete different manner. For instance, a classical bit
can be copied as many times as needed. On the other hand, a quantum
bit cannot be duplicated, due to the well-known {\it no cloning
  property} of quantum states~\cite{nielsen02,preskill99}.  However,
quantum data types are very powerful, due to the phenomena of quantum
superposition and entanglement.

The semantics described in this paper is operational; a program is an
abstract machine with reductions rules. The reduction rules are
probabilistic.  

Some care is needed when defining a type system for higher-order
quantum functions. This is because the question of whether a function
is duplicable or not cannot be directly seen from the types of its
arguments or of its value, but rather it depends on the types of any
free variables occurring in the function definition. As it turns out,
the appropriate type system for higher-order quantum functions in our
setting is affine intuitionistic linear logic.

We also address the question of finding a type inference algorithm.
Using the remark that a linear type is a decoration of an
intuitionistic one, we show that the question of deciding whether or
not a program is valid can be reduced to the question of finding an
intuitionistic type for it and to explore a finite number of linear
decorations for the type.

This work is based on the second author's Master's thesis
{\cite{valiron04}}.


\section{Quantum computing basics}

We briefly recall the basic definitions of quantum computing; please
see \cite{nielsen02,preskill99} for a complete introduction to the
subject.  The basic unit of information in quantum computation is a
quantum bit or \define{qubit}. The state of a single qubit is a a
normalized vector of the $2$-dimensional Hilbert space $\mathbb{C}^2$.
We denote the standard basis of $\mathbb{C}^2$ as $\{\ket{0},
\ket{1}\}$, so that the general state of a single qubit can be written
as $\alpha\ket{0}+\beta\ket{1}$, where
$\abs{\alpha}^2+\abs{\beta}^2=1$.  

The state of $n$ qubits is a normalized vector in $\otimes_{i=1}^{n}
\mathbb{C}^2\iso\mathbb{C}^{2^n}$. We write
$\ket{xy}=\ket{x}\otimes\ket{y}$, so that a standard basis vector of
$\mathbb{C}^{2^n}$ can be denoted $\ket{\binrep{i}{n}}$, where
$\binrep{i}{n}$ is the binary representation of $i$ in $n$ digits, for
$0\leq i<2^n$. As a special case, if $n=0$, we denote the unique
standard basis vector in $\mathbb{C}^1$ by $\ket{}$.

The basic operations on quantum states are unitary operations and
measurements. A unitary operation maps an $n$-qubit state to an
$n$-qubit state, and is given by a unitary $2^n\times 2^n$-matrix.  It
is common to assume that the computational model provides a certain
set of built-in unitary operations, including for example the
\define{Hadamard gate} $H$ and the \define{controlled not-gate} ${\it
  CNOT}$, among others:
\[H = \frac{1}{\sqrt{2}}\left(\begin{array}{cc}
1 & 1\\
1 & -1
\end{array}\right),
\hspace{1.5cm}
{\it CNOT} = \left(\begin{array}{cccc}
1 & 0 & 0 & 0\\
0 & 1 & 0 & 0\\
0 & 0 & 0 & 1\\
0 & 0 & 1 & 0\\
\end{array}\right).\]

The measurement acts as a projection. When a qubit
$\alpha\ket{0}+\beta\ket{1}$ is measured, the observed outcome is a
classical bit. The two possible outcomes $0$ and $1$ are observed with
probabilities $\abs{\alpha}^2$ and $\abs{\beta}^2$, respectively.
Moreover, the state of the qubit is affected by the measurement, and
collapses to $\ket{0}$ if $0$ was observed, and to $\ket{1}$ if $1$
was observed.  More generally, given an $n$-qubit state $\ket{\phi}=
\alpha_0\ket{0}\otimes\ket{\psi_0}+\alpha_1\ket{1}\otimes\ket{\psi_1}$,
where $\ket{\psi_0}$ and $\ket{\psi_1}$ are normalized $(n-1)$-qubit
states, then measuring the leftmost qubit results in the answer $i$
with probability $\abs{\alpha_i}^2$, and the resulting state will be
$\ket{i}\otimes\ket{\psi_i}$.


\section{The untyped quantum lambda calculus}


\subsection{Terms}

Our language uses the notation of the intuitionistic lambda calculus.
For a detailed introduction to the lambda calculus, see e.g.
{\cite{barendregt84}}. We start from a standard lambda calculus with
booleans and finite products. We extend this language with three
special quantum operations, which are $\newterm$, $\measureterm$, and
built-in unitary gates. $\newterm$ maps a classical bit to a quantum
bit. $\measureterm$ maps a quantum bit to a classical bit by
performing a measurement operation; this is a probabilistic operation.
Finally, we assume that there is a set $\UnitaryGates{n}$ of built-in
$n$-ary unitary gates for each $n$. We use the letter $U$ to range
over built-in unitary gates. Thus, the syntax of our language is as
follows:
\[
\begin{array}{llcl}
  {\it Term}\quad & M,N,P\quad & ::=\quad & x \bor MN\bor \lambda
  x.M\bor \iftermx{M}{N}{P}\bor 0\bor 1\bor \measureterm\\
  &&&
  |\;\newterm\bor U\bor \produnitterm\bor \prodterm{M,N}\bor
  \letprodterm{x,y}{M}{N}, 
\end{array}
\]
We follow Barendregt's convention for identifying terms up to
$\alpha$-equivalence. We also sometimes use the shorthand notation
$\prodterm{M_1,\ldots, M_n}= \prodterm{M_1, \prodterm{M_2,\ldots}}$.

\subsection{Programs}

The reader will have noticed that we have not provided a syntax for
constant quantum states such as $\alpha\ket{0}+\beta\ket{1}$ in our
language. One may ask why we did not allow the insertion of quantum
states into a lambda term, such as $\lambda
x.(\alpha\ket{0}+\beta\ket{1})$. The reason is that, in the general
case, such a syntax would be insufficient. Consider for instance the
lambda term $(\lambda y.\lambda f.fpy)(q)$, where $p$ and $q$ are
entangled quantum bits in the state
$\ket{pq}=\alpha\ket{00}+\beta\ket{11}$.  Such a state cannot be
represented locally by replacing $p$ and $q$ with some constant qubit
expressions. The non-local nature of quantum states thus forces us to
introduce a level of indirection into the representation of a state of
a quantum program.

\begin{definition}\rm
  A \define{program state} is represented by a triple $[Q,L,M]$, where
  \begin{itemize}
  \item $Q$ is a normalized vector of $\otimes_{i=0}^{n-1}\mathbb{C}^2$,
    for some $n\geq 0$ 
  \item $M$ is a lambda term,
  \item $L$ is a function from $W$ to $\{0,\ldots,n-1\}$, where
    $\FV(M)\subseteq W\subseteq \Variableterms$. $L$ is also called the
    \define{linking function}.
  \end{itemize}
  The set of program states is denoted by $\qstates$.
\end{definition}

The purpose of the linking function is to assign specific free
variables of $M$ to specific quantum bits in $Q$. 
The notion of $\alpha$-equivalence extends naturally to programs, for instance, the states
$[\ket{1}, \{x \mapsto 0\}, \lambda y.x]$ and 
$[\ket{1}, \{z \mapsto 0\}, \lambda y.z]$
are equivalent. The set of program states, up to $\alpha$-equivalence,
is denoted by $\qstates$.

\begin{convention}\label{con-qlm}
In order to simplify the notation, we will often use the following
convention: we use $p_i$ to denote the free variable $x$ such that
$L(x)=i$. A program $[Q,L,M]$ is abbreviated to $[Q, M']$ with
$M' = M[p_{i_1}/x_1]\ldots[p_{i_n}/x_n]$, where $i_k = L(x_k)$.
\end{convention}

\subsection{Linearity}
\label{subsec-linearity}

An important well-formedness property of quantum programs is that
quantum bits should always be {\em uniquely referenced}: roughly, this
means that no two variable occurrences should refer to the same
physical quantum bit. The reason for this restriction is the
well-known no-cloning property of quantum physics, which states that a
quantum bit cannot be duplicated: there exists no physically
meaningful operation which maps an arbitrary quantum bit $\ket{\phi}$
to $\ket{\phi}\otimes\ket{\phi}$. 

Syntactically, the requirement of unique referencing translates into a
{\em linearity condition}: A lambda abstraction $\lambda x.M$ is
called \define{linear} if the variable $x$ is used at most once during
the evaluation of $M$. A well-formed program should be such that
quantum data is only used linearly; however, classical data, such as
ordinary bits, can of course be used non-linearly. Since the decision
of which subterms must be used linearly depends on type information,
we will not formally enforce any linearity constraints until we
discuss a type system in Section~\ref{sec-types}; nevertheless, we
will assume that all our untyped examples are well-formed in the above
sense.

\subsection{Evaluation strategy}

As is usual in defining a programming language, we need to settle on a
reduction strategy. The obvious candidates are call-by-name and
call-by-value. Because of the probabilistic nature of measurement, the
choice of reduction strategy affects the behavior of programs, not
just in terms of efficiency, but in terms of the actual answer
computed. We demonstrate this in an example. Let ${\bf plus}$ be the
boolean addition function, which is definable as $\mathbf{plus} =
\lambda xy.  \iftermx{x}{(\iftermx{y}{0}{1})}{(\iftermx{y}{1}{0})}$.
Consider the term $M=(\lambda x.\mathbf{plus}\;x\;x)(\measureterm (H
(\newterm \;0)))$. 

\paragraph{Call-by-value.} Reducing this in the empty environment,
using the call-by-value reduction strategy, we obtain the following
reductions:
\[ 
\begin{array}{ll}
\CBV & [ |0\rangle, (\lambda x.\mathbf{plus}\;x\;x)(\measureterm (H\;p_0)) ] \\
\CBV & [\frac{1}{\sqrt{2}}(|0\rangle + |1\rangle), 
(\lambda x.\mathbf{plus}\;x\;x)(\measureterm\;p_0) ]\\
\CBV & 
\left\{\begin{array}{l}
[\;\ket{0}, (\lambda x.\mathbf{plus}\;x\;x)(0) ]\\

[\;\ket{1}, (\lambda x.\mathbf{plus}\;x\;x)(1) ]
\end{array}\right.
\CBV
\left\{\begin{array}{l}
[\;\ket{0}, \mathbf{plus}\;0\;0 ]\\

[\;\ket{1}, \mathbf{plus}\;1\;1 ]
\end{array}\right.
\CBV
\left\{\begin{array}{l}
[\;\ket{0}, 0]\\

[\;\ket{1}, 0]
\end{array}\right.
\end{array}
\]
each with a probability of $1/2$. Thus, under call-by-value reduction,
this program produces the boolean value $0$ with probability $1$. Note
that we have used Convention~\ref{con-qlm} for writing these program
states. 

\paragraph{Call-by-name.} 
Reducing the same term under the call-by-name strategy, we obtain in
one step $[\;\ket{}, \mathbf{plus}\;(\measureterm (H (\newterm
\;0)))\;(\measureterm (H (\newterm \;0))))]$, and then with
probability $1/4$, $[\;\ket{01}, 1\;]$, $[\;\ket{10}, 1\;]$,
$[\;\ket{00}, 0\;]$ or $[\;\ket{11}, 0\;]$. Therefore, the boolean
output of this function is $0$ or $1$ with equal probability.

\paragraph{Mixed strategy.}
Moreover, if we mix the two reduction strategies, the program can even
reduce to an ill-formed term. Namely, reducing by call-by-value until
$[ \frac{1}{\sqrt{2}}(|0\rangle + |1\rangle), (\lambda
x.\mathbf{plus}\;x\;x)(\measureterm\;p_0) ]$, and then changing to
call-by-name, we obtain in one step the term $[
\frac{1}{\sqrt{2}}(|0\rangle + |1\rangle),
(\mathbf{plus}\;(\measureterm\;p_0)\;(\measureterm\;p_0) ]$, which is
not a valid program since there are $2$ occurrences of $p_0$.

\vspace{2ex}
\noindent
In the remainder of this paper, we will only consider the
call-by-value reduction strategy, which seems to us to be the most
natural. 


\subsection{Probabilistic reduction systems}\label{term:sec:prob}

In order to formalize the operational semantics of the quantum lambda
calculus, we need to introduce the notion of a probabilistic reduction
system. 

\begin{definition}\rm
  A \define{probabilistic reduction system} is a tuple 
  $(X,U,R,{\it prob})$
  where $X$ is a set of \define{states}, $U\subseteq X$ is a
  subset of \define{value states}, $R\subseteq (X\setminus U)\times X$
  is a set of \define{reductions}, and ${\it prob}:R\rightarrow [0,1]$
  is a \define{probability function}, where $[0,1]$ is the real unit
  interval. Moreover, we impose the following conditions:
  \begin{itemize}
  \item For any $x \in X$, $R_x = \{\; x' \bor  (x,x')\in R \;\}$ is finite.
  \item $\sum_{x'\in R_x} prob(x, x') \leq 1$
  \end{itemize}
\end{definition}

We call $prob$ the one-step reduction, and denote $x \probred{p} y$ to
be $prob(x,y)=p$.
Let us extend $prob$ to the $n$-step reduction
\[ \begin{array}{rcl}
  prob^0(x,y)    & = & \left\{\begin{array}{cl}
                              0 & \textrm{if}\quad x\neq y\\
                              1 & \textrm{if}\quad x = y
                       \end{array}\right.\\
  prob^1(x,y)    & = & \left\{\begin{array}{cl}
                              prob(x,y)& \textrm{if}\quad (x,y)\in R\\
                              0 & \textrm{else}
                       \end{array}\right.\\
  prob^{n+1}(x,y)& = & \sum_{z\in R_x} prob(x,z)prob^n(z,y),
\end{array} \]
and the notation is extended to $x \nprobred{n}{p} y$ to mean $prob^n(x,y)=p$.

We  say  that $y$  is  \define{reachable  in  one step  with  non-zero
probability} from $x$, denoted $x
\probreach_{>0} y$ when $x \probred{p} y$ with $p>0$.  We say that $y$
is \define{reachable with non-zero probability} from $x$, denoted $x
\nprobreach_{>0} y$ when there exists $n$ such that $x \nprobred{n}{p} y$
with $p>0$.

We can then compute the probability to reach $u \in U$ from $x$:
It is a function from $X\times U$ to $\mathbb{R}$ defined by
$prob_U(x,u) = \sum_{n=0}^\infty prob^n(x,u)$.
The total probability for reaching $U$ from $x$ is
$prob_U(x) = \sum_{n=0}^\infty \sum_{u\in U} prob^n(x,u)$.

On the other hand, there is also the probability to \define{diverge}
from $x$, or never reaching anything. This value is
$prob_\infty(x) = \lim_{n\rightarrow\infty} \sum_{y\in X} prob^n(x,y)$.

\begin{lemma}
  For all $x \in X$, $prob_U(x)+prob_\infty(x) \leq 1$.
\end{lemma}

We define the \define{error probability of $x$} to be the number
$prob_{err}(x) = 1-prob_U(x)-prob_\infty(x)$.

\begin{definition}\rm
  We can define a notion of equivalence in $X$:
  \[ x \approx y \quad\textrm{iff}\quad\forall u \in U
  \left\{\begin{array}{l}
           {\it prob}_U(x, u) = prob_U(y, u)\\
           {\it prob}_\infty(x) = {\it prob}_\infty(y)
  \end{array}\right.\]
\end{definition}

\begin{definition}\rm
  In addition to the notion of reachability with non-zero probability,
  there is also a weaker notion of reachability, given by $R$: We will say
  that $y$ is \define{reachable} from $x$ if $x R y$. By the properties
  of ${\it prob}$,
  $x \probreach_{>0} y$ implies $x \Rreach y$
  with $x \Rreach y$ for $xRy$.
  Let us denote by $\nprobreach$ the relation such that
  $x \nRreach y$ iff there exists $n$ such that $x R^n y$,
  with $R^n$ defined as the $n$-th composition of $R$.
  Similarly, $x \nprobreach_{>0} y$ implies $x \nRreach y$.
\end{definition}

\begin{definition}\rm
\label{term:ref:consistent}
In  a probabilistic  reduction system, a state $x$ is called an
\define{error-state} if $x\not\in U$ and
$\sum_{x'\in X} {\it prob}(x,x') < 1$.
An element $x\in X$ is  \define{consistent} if there is no error-state
$e$ such that $x \nRreach e$.
\end{definition}

\begin{lemma}
If $x$ is consistent, then $prob_{err}(x) = 0$. The converse is false.
\end{lemma}

\begin{remark}
  We need the weaker notion of reachability $x\nRreach y$, in addition
  to reachability with non-zero probability $x\probred{>0}^* y$, because a null probability
  of getting a certain result is not an absolute warranty of its
  impossibility. In the QRAM, suppose we have a qubit in state
  $\ket{0}$. Measuring it cannot theoretically yield the value $1$,
  but in practice, this might happen with small probability, due to
  imprecision of the physical operations and decoherence. What will
  happen if we measure this qubit and get $1$? We need to be sure that
  even in this case the program will not crash. Hence we separate in a
  sense the null probability of getting a certain result, and the
  computational impossibility.
\end{remark}


\subsection{Operational semantics}

We will define a probabilistic call-by-value reduction procedure for
the quantum lambda calculus. Note that, although the reduction itself
is probabilistic, the choice of which redex to reduce at each step is
deterministic. 

\begin{definition}\em
  A {\em value} is a term of the following form:
\[
\begin{array}{llcl}
{\it Value}\quad & V,W\quad & ::=\quad & x \bor \lambda
x.M \bor 0 \bor 1\bor \measureterm\bor \newterm \bor U \bor * \bor \prodterm{V,W}.
\end{array}
\]
The set of \define{value states} is $\vstates=\{[Q, L, V]\in\qstates\such
V\in{\it Value}\}$.
\end{definition}

The reduction rules are shown in Table~\ref{table-cbv}, where we have
used Convention~\ref{con-qlm} to shorten the description of states. We
write $[Q,L,M]\probred{p}[Q',L',M']$ for a single-step reduction of
states which takes place with probability $p$.  In the rule for
reducing the term $U\prodterm{p_{j_1},\ldots,p_{j_n}}$, $U$ is an
$n$-ary built-in unitary gate, $j_1,\ldots,j_n$ are pairwise distinct,
and $Q'$ is the quantum state obtained from $Q$ by applying this gate
to qubits $j_1,\ldots,j_n$.  In the rule for measurement, $\ket{Q_0}$
and $\ket{Q_1}$ are normalized states of the form $ \ket{Q_0} = \sum_j
\alpha_j\ket{\phi_{j}^{0}} \otimes\ket{0}\otimes\ket{\psi_{j}^{0}}$
and $\ket{Q_1} = \sum_j
\beta_j\ket{\phi_{j}^{1}}\otimes\ket{1}\otimes\ket{\psi_{j}^{1}}, $
where $\phi^0_j$ and $\phi^1_j$ is an $i$-qubit state (so that the
measured qubit is the one pointed to by $p_i$). In the rule for for
$\newterm$, $Q$ is an $n$-qubit state, so that $Q\otimes\ket{i}$ is an
$(n+1)$-qubit state, and $p_n$ refers to its rightmost qubit.

\begin{table}[t]
\def\mynl{\\\\[-1ex]}
\[
\begin{array}{c}
[Q,(\lambda x.M)V] \probred{1} [Q, M[V/x]]

\mynl

\infer{[Q,MN] \probred{p} [Q', MN']}{[Q,N] \probred{p} [Q', N']}

\mynl

\infer{[Q,MV] \probred{p} [Q', M'V]}{[Q,M] \probred{p} [Q', M']}

\mynl

\infer{[Q,\langle M_1,M_2\rangle] \probred{p} [Q', \langle
    M_1',M_2\rangle]}{[Q,M_1] \probred{p} [Q', M_1']}

\mynl

\infer{[Q,\langle V_1,M_2\rangle] \probred{p} [Q', \langle
    V_1,M_2'\rangle]}{[Q,M_2] \probred{p} [Q', M_2']}

\end{array}
\hspace{0.5cm}
\begin{array}{c}
[Q,\iftermx{0}{M}{N}] \probred{1} [Q, N]

\mynl

[Q,\iftermx{1}{M}{N}] \probred{1} [Q, M]

\mynl

[Q,U\prodterm{p_{j_1},\ldots,p_{j_n}}] \probred{1} [Q', \prodterm{p_{j_1},\ldots,p_{j_n}}]

\mynl

[ \alpha\ket{Q_0} + \beta\ket{Q_1}, \measureterm\; p_i]
  \probred{\abs{\alpha}^2} [\ket{Q_0}, 0 ] 

\mynl

[ \alpha\ket{Q_0} + \beta\ket{Q_1}, \measureterm\; p_i]
  \probred{\abs{\beta}^2} [\ket{Q_1}, 1 ]

\mynl

[Q,\newterm \; 0] \probred{1} [Q\otimes|0\rangle, p_{n}]

\mynl

[Q,\newterm \; 1] \probred{1} [Q\otimes|1\rangle, p_{n}]

\end{array}
\]
\[
\begin{array}{c}

\infer{[Q,\iftermx{P}{M}{N}]\probred{p}[Q',\iftermx{P'}{M}{N}]}
         {[Q,P]\probred{p}[Q',P']}

\mynl
\infer{
[Q, \textrm{let }\langle x_1,x_2\rangle=
M\textrm{ in }N]
\probred{p} 
[Q', \textrm{let }\langle x_1,x_2\rangle=
M'\textrm{ in }N]
}{[Q,M]\probred{p}[Q',M']}

\mynl

[Q, \textrm{let }\langle x_1,x_2\rangle=
\langle V_1,V_2\rangle\textrm{ in }N]
\probred{1} [Q, N[V_1/x_1,V_2/x_2]]
\mynl
\end{array}
\]
\caption{Reductions rules of the quantum lambda calculus}
\label{table-cbv}
\vspace{-5ex}
\end{table}

We define a weaker relation $\rightsquigarrow$. This relation models
the transformations that can happen in the presence of decoherence and
imprecision of physical operations.  We define $[Q,M]\rightsquigarrow
[Q',M']$ to be $[Q,M]\probreach_p [Q',M']$, even when $p=0$, plus the
additional rule, if $Q$ and $Q'$ are in the same vector space: $[Q,M]
\rightsquigarrow [Q',M]$.

\begin{lemma}
  Let $prob$ be the function such that for $x,y\in\qstates$,
  $prob(x,y)=p$ if $x \probred{p} y$ and $0$ else. Then $(\qstates,
  \vstates, \rightsquigarrow, prob)$ is a probabilistic reduction
  system. $\square$
\end{lemma}

Evidently, this probabilistic reduction system has error states, for
example, $[Q,H(\lambda x.x)]$ or $[Q,U\prodterm{p_0,p_0}]$. Such error
states correspond to run-time errors. In the next section, we
introduce a type system designed to rule out such error states.


\section{The typed quantum lambda-calculus}
\label{sec-types}

We will now define a type system designed to eliminate all run-time
errors arising from the reduction system of the previous section. We
need base types (such as $\bittype$ and $\qbittype$), function types,
and product types. In addition, we need the type system to capture a
notion of duplicability, as discussed in
Section~\ref{subsec-linearity}. We follow the notation of linear logic
{\cite{girard87}}. By default, a term of type $A$ is assumed to be
non-duplicable, and duplicable terms are given the type $\bang{A}$
instead.  Formally, the set of types is defined as follows, where
$\alpha$ ranges over a set of type constants and $X$ ranges over a
countable set of type variables:
\[
\begin{array}{llcl}
{\it \qType}\quad & A,B\quad & ::=\quad \alpha 
\; | \; X 
\; | \; \bang{A} 
\; | \; (A \loli B)
\; | \; \produnittype
\; | \; (A \otimes B)
\end{array}
\]
Note that, because all terms are assumed to be non-duplicable by
default, the language has a linear function type $A\loli B$ and a
linear product type $A\otimes B$. This reflects the fact that there is
in general no canonical diagonal function $A \rightarrow A\otimes A$.
Also, $\produnittype$ is the linear unit type. This will be made more
formal in the typing rules below. We write $\nbang{n}{A}$ for
${!!!\ldots!!}\!A$, with $n$ repetitions of $!$. We also write $A^n$
for the $n$-fold tensor product $A\otimes\ldots\otimes A$.


\subsection{Subtyping}

The typing rules will ensure that any value of type $!A$ is
duplicable. However, there is no harm in using it only once; thus,
such a value should also have type $A$. For this reason, we define a
subtyping relation $\subtype$ as follows:
\[
\begin{array}{c}
\infer[{\it (ax)}\ ]{ \alpha \subtype \alpha}{}
\quad
\infer[{\it (var)}\ ]{ X \subtype X }{}
\quad
\infer[(\top)]{ \top \subtype \top}{}
\quad
\infer[(D)\ ]
        { \bang{A} \subtype B }
        {  A \subtype B }
\quad
\infer[(!)\ ]
        { \bang{A} \subtype \bang{B} }
        { \bang{A} \subtype  B }
\\\\
\infer[(\otimes)]{
  A_1\otimes A_2 \subtype
  B_1\otimes B_2
}{
  A_1 \subtype B_1
  &
  A_2 \subtype B_2
}
\quad
\infer[(\loli)\ ]
        { A' \loli B \subtype A \loli B' }
        { 
          A \subtype A'
        &
          B \subtype B'
        }
\end{array}
\]


\begin{lemma} \label{lem:subtype:set1set2} 
  For any types $A$ and $B$, if $A \subtype B$ and $(m=0)\vee(n\geq 1)$,
  then $\nbang{n}{A} \subtype \nbang{m}{B}$.\qed
\end{lemma}
Notice that one can rewrite types using the notation:
\[
\begin{array}{llcl}
{\it \qType}\quad & A,B\quad & ::=\quad \nbang{n}{\alpha}
\; | \; \nbang{n}{X}
\; | \; \nbang{n}{(A \loli B)}
\; | \; \nbang{n}{\produnittype}
\; | \; \nbang{n}{(A \otimes B)}
\end{array}
\]
with $n\in\mathbb{N}$. Using the overall condition on $n$ and $m$ that
$(m=0)\vee(n\geq 1)$, the rules can be re-written as:
\[ 
\begin{array}{c}
  \infer[(var_2)]
        { \nbang{n}{X} \subtype \nbang{m}{X} }
        { }\quad
\infer[(\alpha)]
        { \nbang{n}{\alpha} \subtype \nbang{m}{\alpha} }
        {}\quad
\infer[(\top)]{ \nbang{n}{\top} \subtype \nbang{m}{\top}}{}
\\\\
\infer[(\otimes)]{
  \nbang{n}{(A_1\otimes A_2)} \subtype
  \nbang{m}{(B_1\otimes B_2)}
}{
  A_1 \subtype B_1
  &
  A_2 \subtype B_2
}\quad
\infer[(\loli_2)]
        { \nbang{n}{(A' \loli B)} \subtype \nbang{m}{(A \loli B')} }
        { 
          A \subtype A'
          &
          B \subtype B'
        }
\end{array}
\]
The two sets of rules are equivalent. 

\begin{lemma} \label{lem:subtype:reversible} 
The rules of the second set are reversible.\qed
\end{lemma}

\begin{lemma} \label{lem:qtype:refltrans} 
  $(\qType, \subtype)$ is reflexive and transitive. If we define an
  equivalence relation $\doteqdot$ by $A \doteqdot B$ iff $A \subtype
  B$ and $B \subtype A$, $(\qType/{\doteqdot}, \subtype)$ is a poset.
  \qed
\end{lemma}

\begin{lemma} \label{lem:qtype:bang} 
If $A \subtype \bang{B}$, then there exists $C$ such that $A=\bang{C}$.
\qed
\end{lemma}


\begin{table*}[t]
\def\mynl{\\\\[-1ex]}
\begin{center}
\begin{tabular}{c}
\begin{tabular}{c c}
$$
\infer[({\it ax}_1)]{\Delta, \typ{x}{A} \entail x:B}{A \subtype B}
$$ &$$
\infer[({\it ax}_2)]{\Delta \entail c:B}{A_c \subtype B}
$$
\end{tabular}\mynl

$$
\infer[({\it if})]
      { \Gamma_1, \Gamma_2, !\Delta \entail \iftermx{P}{M}{N} : A}
      {
        \Gamma_1, \bang{\Delta} \entail P : bit
        &
        \Gamma_2, \bang{\Delta} \entail M : A
        &
        \Gamma_2, \bang{\Delta} \entail N : A
      }
$$\mynl

$$
\infer[({\it app})]
      {\Gamma_1, \Gamma_2, \bang{\Delta} \entail M N : B}
      {
        \Gamma_1, \bang{\Delta} \entail M : A \loli B
        &
        \Gamma_2, \bang{\Delta} \entail N : A
      }
$$\mynl
\begin{tabular}{c c}
$$
\infer[(\lambda_1)]
      { \Delta \entail \lambda x.M : A \loli B}
      { \typ{x}{A}, \Delta \entail M : B}
$$& $$
\infer[(\lambda_2)]
      { \Gamma, \bang{\Delta} \entail \lambda x.M : \nbang{n+1}{(A \loli B)}}
      { \begin{array}{@{}c@{}}\mbox{If $FV(M)\cap|\Gamma| = \emptyset$:}\\\Gamma, \bang{\Delta}, \typ{x}{A} \entail M : B
      \end{array}}
      $$\\
\end{tabular}\mynl

\begin{tabular}{c c}
$$
\infer[(\otimes.I)]{
  \bang{\Delta},\Gamma_1,\Gamma_2
  \entail
  \langle M_1,M_2\rangle : \nbang{n}{({A_1}\otimes{A_2})}
}{
  \bang{\Delta},\Gamma_1 \entail M_1 : \nbang{n}{A_1}
  &
  \bang{\Delta},\Gamma_2 \entail M_2 : \nbang{n}{A_2}
}
$$
&
$$
\infer[(\produnittype)]{
  \Delta \entail \produnitterm : \nbang{n}{\produnittype}
}{}
$$
\end{tabular}
\mynl
$$
\infer[(\otimes.E)]{
  \bang{\Delta}, \Gamma_1, \Gamma_2 
  \entail
  \textrm{let }\langle x_1,x_2\rangle=M\textrm{ in }N :A
}{
  \bang{\Delta}, \Gamma_1\entail M:\nbang{n}{(A_1\otimes A_2)}
  &
  \bang{\Delta},
  \Gamma_2,~x_1{:}\nbang{n}{A_1},~x_2{:}\nbang{n}{A_2}\entail N:A
}
$$\\
\end{tabular}
\end{center}
\caption{Typing rules}
\label{type:typrules}
\vspace{-5ex}
\end{table*}

\subsection{Typing rules}

We need to define what it means for a quantum state $[Q,L,M]$ to be
well-typed. It turns out that the typing does not depend on $Q$ and
$L$, but only on $M$. We introduce typing judgments of the form
$\Delta\entail M:B$. Here $M$ is a term, $B$ is a $\qType$, and
$\Delta$ is a typing context, i.e., a function from a set of variables
to $\qType$. As usual, we write $\abs{\Delta}$ for the domain of
$\Delta$, and we denote typing contexts as
$\typ{x_1}{A_1},\ldots,\typ{x_n}{A_n}$. As usual, we write
$\Delta,\typ{x}{A}$ for $\Delta\cup\s{\typ{x}{A}}$ if
$x\not\in\abs{\Delta}$. Also, if
$\Delta=\typ{x_1}{A_1},\ldots,\typ{x_n}{A_n}$, we write
$\bang{\Delta}=\typ{x_1}{\bang{A_1}},\ldots,\typ{x_n}{\bang{A_n}}$. A
typing judgement is called \define{valid} if it can be derived from
the rules in Table~\ref{type:typrules}.

The typing rule $({\it ax})$ assumes that to every constant $c$ of the
language, we have associated a fixed type $A_c$. The types $A_c$ are
defined as follows:
\[\begin{array}{l@{\hspace{1cm}}l@{\hspace{1cm}}c}
A_0 = \bang{\bit} &
A_{\newterm} = \bang{(\bit \loli \qbit)}\\
A_1 = \bang{\bit} &
A_{\measureterm} = \bang{(\qbit \loli \bang{\bit}}) &
A_U = \bang{(\qbit^n\loli\qbit^n)}
\end{array}\]

Note that we have given the type $\bang{(\bit \loli \qbit)}$ to the
term $\newterm$. Another possible choice would have been
$\bang{(\bang{\bit}\loli \qbit)}$, which makes sense because all
classical bits are duplicable. However, since $\bang{(\bit \loli
  \qbit)}\subtype \bang{(\bang{\bit}\loli \qbit)}$, the second type is
less general, and can be inferred by the typing rules.

Note that, if $[Q,L,M]$ is a program state, the term $M$ need not be
closed; however, all of its free variables must be in the domain of
$L$, and thus must be of type $\qbit$. We therefore define:

\begin{definition}\rm
  A program state $[Q,L,M]$ is \define{well-typed of type} $B$ if
  $\Delta\entail M:B$ is derivable, where
  $\Delta=\s{\typ{x}{\qbittype}\such x\in\FV(M)}$. In this case, we
  write $[Q,L,M]:B$. 
\end{definition}

\subsection{Example: quantum teleportation}

Let us illustrate the quantum lambda calculus and the typing rules
with an example. The following is an implementation of the well-known
quantum teleportation protocol (see e.g. {\cite{nielsen02}}).  The
purpose of the teleportation protocol is to send a qubit from location
$A$ to location $B$, using only classical communication and a
pre-existing shared entangled quantum state. In fact, this can be
achieved by communicating only the content of two classical bits. 

In terms of functional programming, the teleportation procedure can be
seen as the creation of two non-duplicable functions $f:\qbittype
\loli \bittype\otimes\bittype$ and $g:\bittype\otimes\bittype \loli
\qbittype$, such that $f\circ g(x)=x$ for an arbitrary qubit $x$.

We start by defining the following functions ${\bf
  EPR}:{!(\produnittype\loli(\qbittype\otimes\qbittype))}$, ${\bf
  BellMeasure}:
{!(\qbittype\loli(\qbittype\loli\bittype\otimes\bittype))}$, and ${\bf
  U:{!(\qbittype\loli(\bittype\otimes\bittype\loli \qbittype))}}$:
\[\begin{array}{lll}
  {\bf EPR} &=& \lambda x. {\it
    CNOT}\prodterm{H(\newterm 0), \newterm 0},
  \\[1ex]
  {\bf BellMeasure} &=& \lambda q_2.\lambda q_1.
  ({\it let}\ \prodterm{x,y} = {\it CNOT}\prodterm{q_1, q_2}
  \ {\it in}\ \prodterm{\measureterm (H x),\measureterm y},
  \\[1ex]
  {\bf U} &=& \lambda q.\lambda\prodterm{x,y}.\mbox{{\it if $x$}}
  \begin{array}[t]{l}
    \mbox{\it then $(\iftermx{y}{U_{11}q}{U_{10}q})$}
    \\ \mbox{\it else $(\iftermx{y}{U_{01}q}{U_{00}q})$},
  \end{array}
\end{array}
\]
where
\[
\begin{array}{llll}
U_{00}=\left(\begin{array}{cc}1&0\\0&1\end{array}\right),&
U_{01}=\left(\begin{array}{cc}0&1\\1&0\end{array}\right),&
U_{10}=\left(\begin{array}{cc}1&0\\0&-1\end{array}\right),&
U_{11}=\left(\begin{array}{cc}0&1\\-1&0\end{array}\right).
\end{array}
\]
The function ${\bf EPR}$ creates an entangled state
$\frac{1}{\sqrt{2}}(\ket{00}+\ket{11})$.  The function ${\bf
  BellMeasure}$ performs a so-called Bell measurement, and the
function ${\bf U}$ performs a unitary correction on the qubit $q$
depending on the value of two classical bits. We can now construct a
pair of functions $f:\qbittype \loli \bittype\otimes\bittype$ and
$g:\bittype\otimes\bittype \loli \qbittype$ with the above property by
the following code:
\[
\begin{array}{l}
\letprodterm{x,y}{{\bf EPR} \produnitterm\\}{
  {\it let}\ {f}\ = {\bf BellMeasure}\ x\\\ \quad{\it in}\ {\it let}\
  {g}\ = {\bf U}\ y.\\\ \quad{\it in}\ \prodterm{f, g}.
}
\end{array}
\]
The functions $f$ and $g$ thus created do indeed have the desired
property that $f\circ g(x)=x$, where $x$ is any qubit. Note that,
since $f$ and $g$ depend on the state of the qubits $x$ and $y$,
respectively, these functions cannot be duplicated, which is reflected
in the fact that the types of $f$ and $g$ do not contain a top-level
``!''.



\subsection{Properties of the type system}

We derive some basic properties of the type system. 

\begin{definition}\rm
  We extend the subtyping relation to contexts by writing $\Delta
  \subtype \Delta'$ if $|\Delta'|=|\Delta|$ and for all $x$ in
  $|\Delta'|$, $\Delta_f(x) \subtype \Delta'_f(x)$.
\end{definition}

\begin{lemma}\label{lem:weakdash}\label{lem:typrule}\label{lem:subred1} 
\begin{enumerate}
\item If $x\not\in{\it FV}(M)$ and $\Delta, x{:}A\entail M{:}B$, then
  $\Delta \entail M{:}B$.
\item If $\Delta \entail M{:}A$, then $\Gamma,
  \Delta \entail M{:}A$.
\item If $\Gamma \subtype \Delta$ and $\Delta \entail N:A$ and $A
  \subtype B$, then $\Gamma \entail N:B$.
\end{enumerate}
\end{lemma}

\void{
\begin{proof}
By structural induction.\qed
\end{proof}
}

The next lemma is crucial in the proof of the substitution lemma. Note
that it is only true for a value $V$, and in general fails for an
arbitrary term $M$. 

\begin{lemma} \label{lem:refl0} 
  If $V$ is a value and $\Delta \entail V:\bang{A}$, then for all
  $x\in\FV(V)$, there exists some $U\in{\it qType}$ such that
  $\Delta(x)=\bang{U}$.
\end{lemma}

{
\begin{proof}
  By induction on $V$. 
  \begin{itemize}
  \item If $V$ is a variable $x$, then the last rule in the derivation
    was $\begin{array}{c}\infer{\Delta', x:B \entail x:\bang{A}}{B
        \subtype \bang{A}}\end{array}$. Since $B\subtype \bang{A} $,
    $B$ must be exponential by Lemma~\ref{lem:qtype:bang}.
  \item If $V$ is a constant $c$, then $\FV(V)=\emptyset$, hence the
    result holds vacuously.
  \item If $V=\lambda x.M$, the only typing rule that applies is
    $(\lambda_2)$, and $\Delta=\Gamma,!\Delta'$ with
    $FV(M)\cap|\Delta'|=\emptyset$. So every $y\in\FV(M)$ except maybe
    $x$ is \bangnoted{}. Since $\FV(\lambda
    x.M)=(\FV(M)\setminus\{x\})$, this suffices. 
  \item The remaining cases are similar. \qed
  \end{itemize}
\end{proof}
}

\begin{lemma}[Substitution] \label{lem:refl} 
  If $V$ is a value such that $ \Gamma_1, !\Delta, \typ{x}{A} \entail
  M:B$ and $\Gamma_2, !\Delta\entail V:A$, then $\Gamma_1, \Gamma_2,
  !\Delta \entail M[V/x]:B$.
\end{lemma}

\void{
\begin{proof}
  By structural induction on the derivation of $ \Gamma_1, !\Delta,
  \typ{x}{A} \entail M:B$.\qed
\end{proof}
}

\begin{corollary}\label{lem:subred3} 
  If $\Gamma_1, \bang{\Delta}, \typ{x}{A} \entail M:B$ and $\Gamma_2,
  !\Delta \entail V:\nbang{n}{A}$, then $\Gamma_1, \Gamma_2, !\Delta
  \entail M[V/x]:B$.
\end{corollary}

{
\begin{proof}
From Lemma~\ref{lem:refl} and Lemma~\ref{lem:subred1}(3).
\end{proof}
}

\subsection{Subject reduction and progress}

\begin{theorem}[Subject Reduction]\label{the:subred} 
  Given $[Q,L,M]:B$ and $[Q,L,M] \nRreach [Q',L',M']$, then
  $[Q',L',M']:B$.
\end{theorem}

{
\begin{proof}
  It suffices to show this for $[Q,L,M] \probred{p} [Q',L',M']$, and
  we proceed by induction on the rules in Table~\ref{table-cbv}.  The
  rule $[Q, (\lambda x.M)V] \probred{1} [Q, M[V/x]]$ and the rule for
  ``let'' use the substitution lemma. The remaining cases are direct
  applications of the induction hypothesis.\qed
\end{proof}
}

\begin{theorem}[Progress]\label{the:subred2} 
  Let $[Q, L, M]:B$ be a well-typed program. Then $[Q, L, M]$ is not
  an error state in the sense of Definition~\ref{term:ref:consistent}.
  In particular, either $[Q, L, M]$ is a value, or else there exist
  some state $[Q', L', M']$ such that $[Q, L, M]\probred{p}[Q', L',
  M']$. Moreover, the total probability of all possible single-step
  reductions from $[Q, L, M]$ is $1$.
\end{theorem}

\begin{corollary}
  Every sequence of reductions of a well-typed program either
  converges to a value, or diverges. 
\end{corollary}

The proof of the Progress Theorem is similar to the usual proof, with
two small differences. The first is the presence of probabilities, and
the second is the fact that $M$ is not necessarily closed. However,
all the free variables of $M$ are of type $\qbit$, and this property
suffices to prove the following lemma, which generalizes the usual
lemma on the shape of closed well-typed values:

\begin{lemma}\label{lem-progress-value}
  Suppose $\Delta=\typ{x_1}{\qbit},\ldots,\typ{x_n}{\qbit}$, and $V$
  is a value. If $\Delta\entail V:A\loli B$, then $V$ is $\newterm$,
  $\measureterm$, $U$, or a lambda abstraction. If $\Delta\entail
  V:A\otimes B$, then $V=\prodterm{V_1,V_2}$. If $\Delta\entail
  V:\bit$, then $V=0$ or $V=1$. \qed
\end{lemma}

\begin{proofof}{the Progress Theorem}
  By induction on $M$. The claim follows immediately in the cases when
  $M$ is a value, or when $M$ is a left-hand-side of one of the rules
  in Table~\ref{table-cbv} that have no hypotheses. Otherwise, using
  Lemma~\ref{lem-progress-value}, $M$ is one of the following: $PN$,
  $NV$, $\prodterm{N,P}$, $\prodterm{V,N}$, $\iftermx{N}{P}{Q}$,
  $\letprodterm{x,y}{N}{P}$, where $N$ is not a value. In this case,
  the free variables of $N$ are still all of type $\qbit$, and by
  induction hypothesis, the term $[Q,L,N]$ has reductions with total
  probability $1$, and the rules in Table~\ref{table-cbv} ensure that
  the same is true for $[Q,L,M]$.\qed
\end{proofof}

\void{
\begin{proofof}{the Progress Theorem}
  By induction on $M$. The term $M$ is of one of the following forms,
  where $N$ is not a value:
  \begin{enumerate}
  \item $x$, $\lambda x.M$, $0$, $1$, $\measureterm$, $\newterm$, $U$,
    $*$, $\prodterm{V,W}$. 
  \item $VW$, $\iftermx{V}{P}{Q}$, $\letprodterm{x,y}{V}{P}$,
  \item $NV$, $PN$, $\iftermx{N}{P}{Q}$, $\prodterm{V,N}$,
    $\prodterm{N,P}$, $\letprodterm{x,y}{N}{P}$,
  \end{enumerate}
  The terms in the first list are values, and there is nothing to
  show. The terms in the second list have the property that $V$ has
  type $A\loli B$, $\bit$, or $A\otimes B$, respectively. By
  Lemma~\ref{lem-progress-value}, $M$ is the left-hand-side of one of
  the rules in Table~\ref{table-cbv} with no hypotheses. The terms in
  the second list have the property that the free variables of $N$ are
  all of type $\qbit$, thus the induction hypothesis applies to them
  and therefore $[Q,L,N]$ has reductions with total probability
  $1$. By the rules in Table~\ref{table-cbv}, the same is true for
  $M$. 
\end{proofof}
}

\section{Type inference algorithm}

It is well-known that in the simply-typed lambda calculus, as well as
in many programming languages, satisfy the \define{principal type
  property}: every untyped expression has a most general type,
provided that it has any type at all. Since most principal types can
usually be determined automatically, the programmer can be relieved
from the need to write any types at all. 

In the context of our quantum lambda calculus, it would be nice to
have a type inference algorithm; however, the principal type property
fails due to the presence of exponentials $!A$. Not only can an
expression have several different types, but in general none of the
types is ``most general''. For example, the term $M=\lambda xy.xy$ has
possible types $T_1={(A\loli B)}\loli{(A\loli B)}$ and
$T_2=\bang{(A\loli B)}\loli\bang{(A\loli B)}$, among others. Neither
of $T_1$ and $T_2$ is a substitution instance of the other, and in
fact the most general type subsuming $T_1$ and $T_2$ is $X\loli X$,
which is not a valid type for $M$. Also, neither of $T_1$ and $T_2$ is
a subtype of the other, and the most general type of which they are
both subtypes is ${(A\loli B)}\loli\bang{(A\loli B)}$, which is not a
valid type for $M$. 

In the absence of the principal type property, we need to design a
type inference algorithm based on a different idea. The approach we
follow is the one suggested by V. Danos, J.-B. Joinet and H.
Schellinx~\cite{danos95}. The basic idea is to view a linear type as a
``decoration'' of an intuitionistic type. Our type inference algorithm
is based on the following technical fact, given below: if a given
term has an intuitionistic type derivation $\pi$, then it is
linearly typable if and only if there exists a linear type derivation
which is a decoration of $\pi$. Typability can therefore be decided
by first doing intuitionistic type inference, and then checking
finitely many possible linear decorations.

\subsection{Skeletons and decorations}

The class of {\em intuitionistic types} is
\[
\begin{array}{llcl}
{\it iType}\quad & U,V\quad & ::=\quad & \alpha \;|\;X\;|\;(U\duploli
V)\;|\;(U\times V)\;|\;\top
\end{array}
\]
where $\alpha$ ranges over the type constants and $X$ over the type
variables.

To each $A\in\qType$, we associate its \define{type skeleton}
$\skel{A}\in\iType$, which is obtained by removing all occurrences of
``$!$''.  Conversely, every $U\in\iType$ can be lifted to some
$\invskel{U}\in\qType$ with no occurrences of ``$!$''. Formally:

\begin{definition}\rm
  The functions $\dagger:\qType\ii\iType$ and
  $\clubsuit:\iType\ii\qType$ are defined by:
  \[
  \begin{array}{cc}
    \begin{array}{rcl}
      \skel{\nbang{n}{\alpha}} & = & \alpha\\
      \skel{\nbang{n}{X}} & = & X\\
      \skel{\nbang{n}{(A \loli B)}} & = & \skel{A} \duploli \skel{B}\\
      \skel{\nbang{n}{(A \otimes B)}} & = & \skel{A} \times \skel{B}\\
      \skel{\nbang{n}{\top}} & = & \top.
    \end{array}
    &\quad\quad
    \begin{array}{rcl}
      \invskel{\alpha} & = & \alpha
      \\
      \invskel{X} & = & X
      \\
      \invskel{(U\duploli V)} & = & \invskel{U}\loli\invskel{V}
      \\
      \invskel{(U\times V)} & = & \invskel{U}\otimes\invskel{V}
      \\
      \invskel{\top} & = & \top
    \end{array}
  \end{array}
  \]  
\end{definition}

\begin{lemma}
  If $A \subtype B$, then $\skel{A}=\skel{B}$.
  If $U\in\iType$, then $U=\skel{\invskel{U}}$.
\end{lemma}
Writing $\Delta\skelentail M:U$ for a typing judgement of the
simply-typed lambda calculus, we can extend the notion of skeleton to
contexts, typing judgments, and derivations as follows:
\[
\begin{array}{rcl}
\skel{\{x_1{:}A_1,\ldots,x_n{:}A_n\}} &=&  
\{x_1{:}\skel{A_1},\ldots,x_n{:}\skel{A_n}\}
\\
\skel{(\Delta\entail M:A)} &=&
(\skel{\Delta} \skelentail M:\skel{A}).
\end{array}
\]
From the rules in Table~\ref{type:typrules}, it is immediate that if
$\Delta\entail M:A$ is a valid typing judgment in the quantum
lambda-calculus, then $\skel{(\Delta\entail
  M:A)}=(\skel{\Delta}\skelentail M:\skel{A})$ is a valid typing
judgment in the simply-typed lambda-calculus.

We now turn to the question of how an intuitionistic typing derivation
can be ``decorated'' with exponentials to yield a valid quantum typing
derivation. These decorations are going to be the heart of the quantum
type inference algorithm.

\begin{definition}\rm
  Given $A\in{\it qType}$ and  $U\in{\it iType}$, we define the
  \define{decoration $U\skeltotype A \in{\it  qType}$ of $U$ along
    $A$} by
  \begin{enumerate}
  \item $U \skeltotype \nbang{n}{A} = \nbang{n}{(U\skeltotype
      A)}$,
  \item $(U\duploli V)\skeltotype(A\loli B) = (U\skeltotype A\loli
    V\skeltotype B)$,
  \item $(U\times V)\skeltotype(A\otimes B) = (U\skeltotype A\otimes
    V\skeltotype B)$, and in all other cases:
  \item $U \skeltotype A = \invskel{U}$.
  \end{enumerate}
\end{definition}

The following lemma is the key to the quantum type inference algorithm:

\begin{lemma} \label{lem:inf:cap}
  If $M$ is well-typed in the quantum lambda-calculus with typing
  judgment $\Gamma \entail M:A$, then for any valid typing judgment
  $\Delta\skelentail M:U$ in simply-typed lambda-calculus with
  $|\Delta|=|\Gamma|$, the typing judgment $\Delta\skeltotype\Gamma
  \entail M:U\skeltotype A$ is valid in the quantum lambda-calculus
  and admits a derivation which has for skeleton the derivation of
  $\Delta\skelentail M:U$.
\end{lemma}

\subsection{Elimination of repeated exponentials}

The type system in Section~\ref{sec-types} allows types with repeated
exponentials such as $!!A$. While this is useful for compositionality,
it is not very convenient for type inference. We therefore consider a
reformulation of the typing rules which only requires single
exponentials.

\void{
\begin{definition}\rm
  For $A\in\qType$, we define $\uniq{A}\in\qType$ to be the result of
  erasing multiple exponentials in $A$. Formally, 
  \[  
  \begin{array}{lll}
    \uniq{\nbang{n+1}{A}} & = & !\uniq{A}, \mbox{if $A$ is not exponential}\\
    \uniq{{\alpha}} & = & \alpha\\
    \uniq{{X}} & = & X\\
    \uniq{{\top}} & = & \top \\
    \uniq{{(A \loli B)}} & = & \uniq{A} \loli \uniq{B}\\
    \uniq{{(A \otimes B)}} & = & \uniq{A} \otimes \uniq{B}.\\
  \end{array}
  \]
  We also extend this operation to typing contexts and judgments in
  the obvious way.
\end{definition}
}

\begin{lemma}
  The following are derived rules of the type system in
  Table~\ref{type:typrules}, for all $\tau,\sigma\in\s{0,1}$.
  \vspace{-2ex}
  \[
  \begin{array}{c}
    \infer[(\otimes.I')]{
      \bang{\Delta},\Gamma_1,\Gamma_2
      \entail
      \langle M_1,M_2\rangle : \bang{(\nbang{\tau}{A_1}\otimes\nbang{\sigma}{A_2})}
      }{
      \bang{\Delta},\Gamma_1 \entail M_1 : \bang{A_1}
      &
      \bang{\Delta},\Gamma_2 \entail M_2 : \bang{A_2}
      }
    \\\\[-1ex]
    \infer[(\otimes.E')]{
      \bang{\Delta}, \Gamma_1, \Gamma_2 
      \entail
      \textrm{let }\langle x_1,x_2\rangle=M\textrm{ in }N :A
      }{
      \bang{\Delta}, \Gamma_1\entail M:\bang{(\nbang{\tau}{A_1}\otimes \nbang{\sigma}{A_2})}
      &
      \bang{\Delta},
      \Gamma_2,~x_1{:}\bang{A_1},~x_2{:}\bang{A_2}\entail N:A
      }
  \end{array}
  \]
\end{lemma}

\begin{lemma}\label{lem-uniq}
  If $M$ is typable in the quantum lambda calculus by some derivation
  $\pi$, then $M$ is typable in the system with the added rules
  $(\otimes.I')$ and $(\otimes.E')$, by a derivation $\pi'$ using no
  repeated exponentials. Moreover, $\skel{\pi'}=\skel{\pi}$.  \qed
\end{lemma}

\subsection{Description of the type inference algorithm}

To decide the typability of a given term $M$, first note the
following: if $M$ is not typable in simply-typed lambda calculus, then
$M$ is not quantum typable. On the other hand, if $M$ admits a typing
judgment $\Gamma\skelentail M:U$ in the simply-typed lambda calculus,
say with typing derivation $\pi$, then by Lemma~\ref{lem:inf:cap}, $M$
is quantum typable if and only if $M$ has a quantum derivation whose
skeleton is $\pi$. Thus we can perform type inference in the quantum
lambda-calculus in two steps:
\begin{enumerate}
\item Find an intuitionistic typing derivation $\pi$, if any.
\item Find a decoration of $\pi$ which is a valid quantum typing
  derivation, if any.
\end{enumerate}
Step $(1)$ is known to be decidable. For step $(2)$, note that by
Lemma~\ref{lem-uniq}, it suffices to consider decorations of $\pi$
without repeated exponentials. Since there are only finitely many such
decorations, the typability of $M$ is clearly a decidable problem.
Also note that if the algorithm succeeds, then it returns a possible
type for $M$.  However, it does not return a description of all
possible types.

It should further be noted that the space of all decorations of $\pi$,
while exponential in size, can be searched efficiently by solving a
system of constraints.

\vspace{-1.5ex}

\section{Conclusion and further work}

\vspace{-.5ex}

In this paper, we have defined a higher-order quantum programming
language based on a linear typed lambda calculus. Compared to the
quantum lambda calculus of van Tonder {\cite{tonder03,tonder04}}, our
language is characterized by the fact that it contains classical as
well as quantum features; for instance, we provide classical datatypes
and measurements as a primitive feature of our language. Moreover, we
provide a subject reduction result and a type inference algorithm. As
the language shows, linearity constraints do not just exist at base
types, but also at higher types, due to the fact that higher-order
function are represented as closures which may in turns contain
embedded quantum data. We have shown that affine intuitionistic linear
logic provides precisely the right type system to deal with this
situation.

There are many open problems for further work. An interesting question
is whether the syntax of this language can be extended to include
recursion. Another question is to study extensions of the type system,
for instance with additive types as in linear logic.  One may also
study alternative reduction strategies. In this paper, we have only
considered the call-by-value case; it would be interesting to see if
there is a call-by-name equivalent of this language.  Finally, another
important open problem is to find a good denotational semantics for a
higher order quantum programming language. One approach for finding
such a semantics is to extend the framework of
Selinger~\cite{selinger04} and to identify an appropriate higher-order
version of the notion of a superoperator.

\appendix

\addcontentsline{toc}{section}{Bibliography}

\bibliographystyle{plain}
\bibliography{project}

\end{document}